\documentclass[12pt]{article}
\usepackage{epsfig}
\usepackage{amsmath}
\usepackage{hhline}
\usepackage{amssymb}
\usepackage{times}

\newlength{\dinwidth}
\newlength{\dinmargin}
\begin{document}  
\def\ddiff{{\rm d}}
\def\xgamma{$x_{\gamma}$}
\def\qsq{$Q^2$}
\def\gev{$\rm ~GeV \ $ }
%
%
%
%
\def\Journal#1#2#3#4{{#1} {\bf #2} (#3) #4}
\def\NCA{\em Nuovo Cimento}
\def\NIM{\em Nucl. Instrum. Methods}
\def\NIMA{{\em Nucl. Instrum. Methods} {\bf A}}
\def\NPB{{\em Nucl. Phys.}   {\bf B}}
\def\PLB{{\em Phys. Lett.}   {\bf B}}
\def\PRL{\em Phys. Rev. Lett.}
\def\PRD{{\em Phys. Rev.}    {\bf D}}
\def\ZPC{{\em Z. Phys.}      {\bf C}}
\def\EUR{{\em Eur. Phys. J.} {\bf C}}
\def\CPC{\em Comp. Phys. Commun.}
\bibliographystyle{st}    
%
\begin{titlepage}
\begin{flushleft}
  {\tt DESY 00-035} \hfill {\tt ISSN 0418-9833} \\
  {\tt March 2000}  
\end{flushleft}

\vspace*{3cm}
\begin{center}
\begin{LARGE}
{\bf Measurement of   Di-jet Cross-Sections  in
  Photoproduction  and Photon Structure} \\
\vspace*{2.5cm}
H1 Collaboration \\
\vspace*{2.5cm}
\end{LARGE}
{\bf Abstract}
\begin{quotation}
\noindent 
 The production of hard di-jet events in photoproduction at HERA is dominated
 by resolved photon processes in which  a parton in the photon with momentum
 fraction $x_{\gamma}$ is scattered from  a
 parton in the proton.  These processes are  sensitive to the quark
 and gluon
 content of the photon.
 The differential di-jet cross-section  $ \ddiff\sigma /
 \ddiff \log(x_{\gamma})$ is presented here,
  measured in tagged photoproduction at HERA using data taken with the H1
 detector, corresponding to  an integrated
 luminosity of 7.2 pb$^{-1}$. 
 Using a restricted data sample at high transverse jet energy, $E_{T,jet} > 6 $
 GeV, 
  the effective parton density
 $f_{\gamma ,eff}(x_{\gamma}) = \left[q(x_{\gamma}) + \bar{q}(x_{\gamma}) +9/4 \ g(x_{\gamma}) \right]$
 in the
 photon in leading order QCD is  measured down to \xgamma = 0.05 from which
 the gluon density in the photon is derived.
\\
\end{quotation}
\vspace*{2.0cm}
\end{center}
\begin{center}
Submitted to Physics Letters
\end{center}
\end{titlepage}
\normalsize
%
%
 
\begin{flushleft}
 C.~Adloff$^{33}$,                
 V.~Andreev$^{24}$,               
 B.~Andrieu$^{27}$,               
 V.~Arkadov$^{35}$,               
 A.~Astvatsatourov$^{35}$,        
 I.~Ayyaz$^{28}$,                 
 A.~Babaev$^{23}$,                
 J.~B\"ahr$^{35}$,                
 P.~Baranov$^{24}$,               
 E.~Barrelet$^{28}$
% ...

\end{flushleft}
 
 
\begin{flushleft} 
  {\it 
 $ ^1$ I. Physikalisches Institut der RWTH, Aachen, Germany$^a$ \\
 $ ^2$ III. Physikalisches Institut der RWTH, Aachen, Germany$^a$ \\
% ...
$ ^{23}$ Institute for Theoretical and Experimental Physics,
          Moscow, Russia \\
 $ ^{24}$ Lebedev Physical Institute, Moscow, Russia$^{e,h}$ \\
% ...
$ ^{27}$ LPNHE, \'{E}cole Polytechnique, IN2P3-CNRS, Palaiseau, France \\
 $ ^{28}$ LPNHE, Universit\'{e}s Paris VI and VII, IN2P3-CNRS,
          Paris, France \\
% ...
$ ^{33}$ Fachbereich Physik, Bergische Universit\"at Gesamthochschule
          Wuppertal, Wuppertal, Germany$^a$ \\
% ...
$ ^{35}$ DESY, Zeuthen, Germany$^a$ \\
% ...
$ ^a$ Supported by the Bundesministerium f\"ur Bildung, Wissenschaft,
        Forschung und Technologie, FRG,
        under contract numbers 7AC17P, 7AC47P, 7DO55P, 7HH17I, 7HH27P,
        7HD17P, 7HD27P, 7KI17I, 6MP17I and 7WT87P \\
% ...
$ ^e$ Supported by the Deutsche Forschungsgemeinschaft \\
% ...
$ ^h$ Supported by Russian Foundation for Basic Research 
       grant no. 96-02-00019 \\
% ...
 }
\end{flushleft}
 
\newpage

\section{Introduction}
The interaction of electrons and protons at the  HERA collider  is dominated
by photoproduction processes in which quasireal photons emitted by the 
electrons interact with the protons. 
The center of mass
energies  in the $\gamma$p system extend to 300 GeV. A fraction of these
events has large transverse energy in the final state and contains  jets.
Previous studies of hard $\gamma $p scattering processes  by   
H1 \cite{h1_tracks,h1_old,h1.jet98} and  ZEUS \cite{zeus_old} 
have shown that the photoproduction of jets can be described
 in perturbative QCD. 
The photon interacts  either directly with a parton from the proton, or
it develops hadronic structure and one of its own partons interacts
with one of those from the proton.  The former  are referred to as direct
 interactions, whereas the
latter are  referred to as resolved interactions. \par
Jet cross-section predictions  are obtained to leading order (LO) 
 in the strong coupling constant   as a convolution of the   hard scattering
cross-sections calculated at the tree level
with 
the parton densities in the photon and  the proton. The partons leaving 
the hard
scattering reaction are identified with jets. In the kinematic
 range of the present analysis the parton
densities in the proton are rather well known and the quark densities of the photon
have  been determined  from two-photon processes at $e^+e^-$ colliders.
 The measurement of the di-jet cross-section can  therefore be used to
determine  the gluon distribution in the photon.
\par
From previous studies \cite{h1_old,jet_hera} it is known  that the energy
 flow in the events
of interest here is complicated by a large ``underlying event'' energy, which
can be described as arising from multiple interactions. That is, in addition
to the primary hard interaction, further interactions occur between partons in
the proton and photon remnants. Modelling of higher order QCD effects using
parton showers is also important if an accurate description of the energy flow
in and around the jets is to be obtained.  Current Monte Carlo (MC) models
 including such  effects
are based on LO QCD matrix element calculations; NLO predictions are available
at the parton level only.
\par
The analysis described in this paper is similar to that presented in
 an earlier publication \cite{h1_old}. However, the data used now
 correspond to an
  integrated luminosity of  7.2 pb$^{-1} $ as opposed  to 0.29 pb$^{-1}$.
 Its main emphasis is on the  study of di-jet production at small \xgamma \ where
 gluons in the photon are expected to make the largest contribution to the
 cross-section. The  data in this kinematic region are  strongly
 affected by
 non-perturbative effects as discussed in detail below.  We therefore  limit
 ourselves here to a LO QCD analysis of the parton
 distributions in the photon. A NLO analysis of di-jet events in
 photoproduction has been published recently by the ZEUS collaboration
 \cite{ZEUSNLO} for a high cut in transverse jet energy $E_T > 11 $ GeV.
In this kinematic region of large $x_{\gamma}$, where the influence of the 
underlying  event energy is reduced, the quark rather than the gluon content
of the photon is expected to dominate the cross-section.

\section{The H1 Detector}

A detailed description of the H1 detector can be found elsewhere \cite{h1detector}.
Here we describe only those components which are important for this analysis.\par
The H1 central tracking system is mounted coaxially around the beam-line
and covers polar angles  $\theta $, measured with respect to the proton beam
direction, in the range 
  $20^\circ  < \theta < 160^\circ$.
 Momentum measurements of charged particles are provided by two 
cylindrical drift chambers. The
central tracking system is complemented at two radii by $z$- drift chambers,
which provide accurate
 measurements of
the $z$ coordinate along the beam line of charged particle tracks,
 and  multiwire proportional
chambers (MWPCs), which allow   triggering  on central  tracks.
 In the present analysis 
the tracking detectors are used to define the vertex position along the
beam axis and to improve  the measurement of the hadronic energy flow at low
hadron energies.

The tracking system is surrounded by a highly segmented liquid argon (LAr) 
sampling calorimeter  with an inner electromagnetic section consisting 
of lead absorber plates with a total depth of 20 to 30 radiation lengths and
an outer hadronic section with steel absorber plates.
The LAr calorimeter  covers polar angles between $4^\circ$ and $154^\circ$ 
with full azimuthal acceptance. The total depth of the calorimeter varies
between 4.5 and 8 hadronic interaction lengths. 
The energy resolution was measured to be  $\sigma (E)/E \approx 0.12/\sqrt{E}$ for electrons and
$\sigma (E)/E \sim 0.5/\sqrt{E}$ for hadrons ($E$ in GeV) in test beam
experiments. The absolute energy
scale is known for the present data sample to a precision of 1 to 3\% for
positrons  and 4\% for hadrons.
The region $153^\circ < \theta < 177.8^\circ$ is covered by a
lead/scintillating-fibre calorimeter.

The luminosity determination is based on the measurement of the 
 bremsstrahlung process,  $ep \rightarrow ep\gamma$,  using the small angle  photon detector
 ($z=-103$ \ m), and by detecting the scattered positron 
 in the small angle electron detector ( $z=-33$ \  m) 
where $z$ is the coordinate along the beam line with the nominal vertex at the
 origin. Both detectors  are
crystal \v{C}erenkov
 calorimeters with an energy resolution of $\sigma (E)/E \approx
 0.22/\sqrt{E}$. The small angle electron detector is used in the present
 analysis also to tag photoproduction events.

\section{Event Selection and Kinematic Reconstruction}

The events used in this analysis  were taken during the 1996 running period,
in which HERA collided 820 GeV protons with 27.5 GeV positrons.

The transverse jet energies required in this analysis are as low as 4 GeV.
In order to
improve the jet energy  resolution at low jet energies the energy flow is  reconstructed by
combining the energy measurements made in the LAr calorimeter with the measured momenta
of spatially associated  charged
tracks with transverse momenta smaller than
1.5   GeV, avoiding double counting. More details are given in
 \cite{highq**2}.

Events were selected according to  the following requirements:
\begin{enumerate}
\item The event is  triggered by a combination of trigger signals
 from the small angle electron detector and from charged tracks in the central detectors with a
 minimum requirement on their transverse momentum of about 300 MeV.
\item The scattered positron is detected and  measured in the small angle
  electron detector
 in order to
ensure a low photon virtuality ($Q^2 < \ 0.01 $ GeV$^2$).
The energy fraction $y_e$ carried by the radiated photon is restricted to
the range $0.5 < y_e < 0.7$, where $y_e$ is reconstructed from the
  energy
of the scattered positron.  
 The lower cut on $y_e$  ensures that a high
 momentum photon enters the 
hard scattering process such that the detector acceptance for the two hard
jets is large; the upper cut is required by the  acceptance of the small angle
electron detector. 

\item At least two jets with transverse energy $E_{T,jet} > 4 $ GeV and an
invariant jet-jet mass $M_{1,2}>12 $  GeV have to be
 found using  a cone algorithm \cite{cdfcone} in the region $-0.5 <
 \eta_{jet} < 2.5$. Here $\eta_{jet}$ is the pseudorapidity in the laboratory
 and positive $\eta $ corresponds to the direction of the outgoing proton. A
 small cone radius of $R=0.7$ in the $\eta - \phi$ plane is used to
 reduce the effects of the underlying event energy on the jet energy
 measurement. The two jets with highest transverse energy are associated to
 the hard scattering process.

\item The difference in pseudorapidity  between the jets is restricted to 
  $|\eta_{jet1}-\eta_{jet2} | < 1$. This cut  reduces the background of events
where one  jet is in the beam pipe and a second  jet, not associated 
 with the hard scattering process, is found instead. 
\end{enumerate}
 Using these  cuts 1889 di-jet events remain.
The longitudinal momentum fraction of the incident parton in the photon is 
estimated using $y_e$ and the transverse 
energies and pseudorapidities of the two jets with the highest $E_T$,
\begin{equation} x_{\gamma,jets} = \frac{E_{T,jet1} e^{-\eta_{jet1}} + 
                            E_{T,jet2} e^{-\eta_{jet2}}}
                           {2 \, y_e \, E_{e,0}}, 
\end{equation}
where $E_{e,0}$ denotes the electron beam energy. 
In the selected event sample, $x_{\gamma}$ is limited to the range 
 $ x_{\gamma,jets} > 0.03$ as a result of the 
 cuts on the transverse energy, on the invariant mass of the jets, 
on the pseudorapidity and on $y_e$.

The trigger efficiency is monitored in the data by using an independent calorimetric
reference trigger. The efficiency ranges from $90 \%$ at 
high $x_{\gamma,jets}$ \  to $65 \%$ at low $x_{\gamma,jets}$,  and is well described by the 
detector simulation.
An error of $\pm 5$\% is assigned to the trigger efficiency.

\section{Monte Carlo Generators for Hard \boldmath $\gamma p$  Processes}

The analysis uses simulated events  to correct the measurements
for detector effects,  and  to further compare the data with perturbative 
 QCD predictions
for the hard parton scattering and different models for multiple interactions.  
The Monte Carlo generators used in this analysis are PHOJET \cite{phojet} 
and  PYTHIA \cite{pythia}.
 Both  use LO QCD matrix elements for the
hard scattering subprocesses. Initial and final state parton radiation and the
string fragmentation model are included as implemented
in the JETSET program  \cite{jetset}. The two Monte Carlo generators differ in the treatment of
multiple interactions and the transition from hard to soft processes at low
transverse parton momentum $\hat p_T$. The hard parton-parton 
cross-section diverges towards low $\hat{p}_t$  and therefore  needs a
regularisation
to normalise to the measured total cross-section. This regularisation is
achieved for PHOJET by a simple cut-off at $\hat{p_t} =2.5 $ GeV. 
For the PYTHIA generator we have chosen the option to use a damping factor
$\hat{p}_t^2/(\hat{p}_t^2 + \hat{p}_{0t}^2)$ where  $ \hat{p}_{0t}$ was taken to
be 1.55 GeV. \footnote{ This
regularisation corresponds to a model with variable impact parameter for
multiple parton interactions as explained in reference \cite{pythia}, 
section 11.2.}

The PHOJET  event generator simulates in a consistent way all components
that contribute to the total photoproduction cross-section.
PHOJET incorporates detailed simulations of both multiple soft and hard parton
interactions on the basis of a unitarisation scheme.

The PYTHIA 5.7 event generator uses LO QCD  calculations to simulate both the
primary
parton-parton scattering process and   
multiple parton interactions.  The latter are considered to result  from the
scattering of partons  from the
photon and proton remnants. The final state partons are required to have a
transverse momentum of at least  $1.2 \ $ GeV in all cases.

For both Monte Carlo models the factorisation and renormalisation scales
were set to the transverse momentum $\hat{p}_t$  of  the scattered  partons.
GRV92-LO    \cite{grv} parton distribution functions for the 
proton and photon were used for the generation of the events.

\section{Energy Flow and Jet Correlations}
A precise measurement 
 of the transverse jet energy $E_{T,jet}$ is  very important 
 because the measured transverse jet energy distribution falls roughly as
 $(E_{T,jet})^{-5.5}$. Therefore a poor  description of the energy flow around the
jet  leads to severe systematic biases in the determination of cross-sections. 

\begin{figure}[ht]
\epsfig{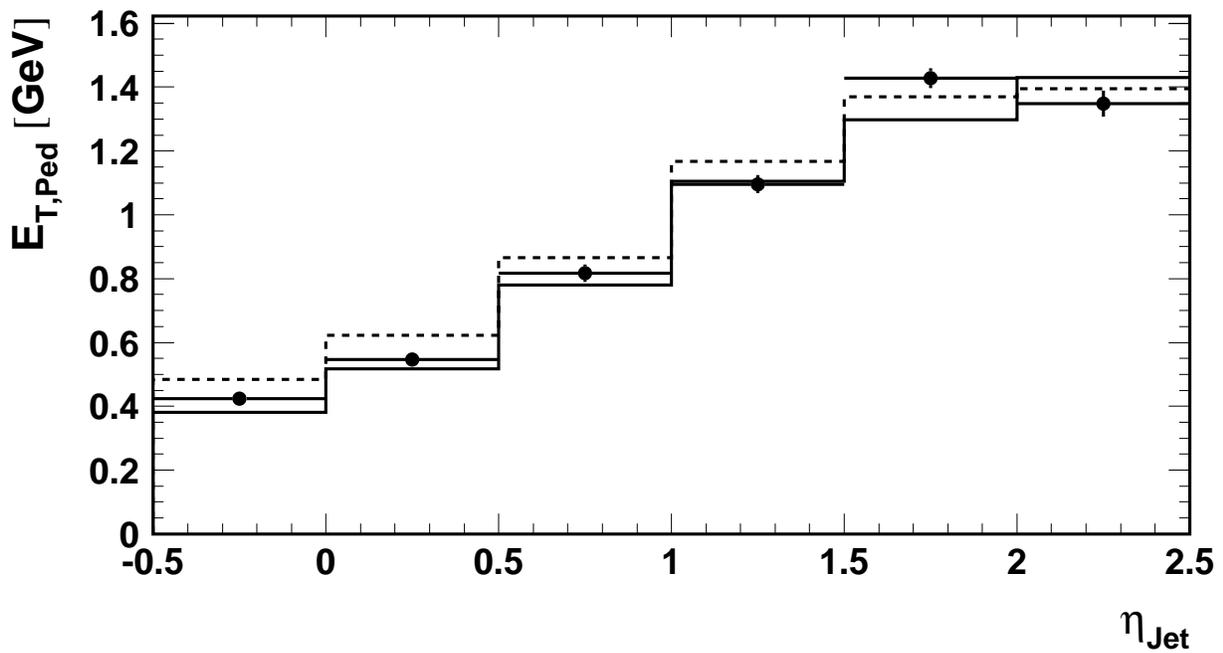}
\caption{Mean transverse energy density (per unit area in $\eta -
  \phi$) outside of the
   jets (pedestal energy) 
evaluated in the interval $-1 < \eta -\eta_{jet} <1$ as a function of jet  
pseudorapidity. The measured distribution at detector level (full circles)
is compared to  the predictions of PYTHIA  (dashed line) and
 PHOJET  (full line).} 

\label{energyflow}
\end{figure}

 The transverse energy flow
is well described by both Monte Carlo simulations within the jet cone. 
They  differ, however, 
outside the jets: 
  PYTHIA slightly overestimates and 
PHOJET underestimates the transverse energy \cite{kaufmann}.
 Remaining differences
between the two Monte Carlo models 
are  used to estimate the
systematic error of the jet reconstruction due to the underlying event energy.
 The transverse energy outside
the jet cones  depends mainly  on $\eta $ because the energy available for multiple 
interactions
is large for small \xgamma , i.e. large $\eta$, where the photon spectator has
large fractional momentum $1-x_{\gamma} $. The average transverse energy density (per unit area in
$\eta - \phi $) outside  the jets in the interval
 $-1 < \eta - \eta_{jet} < 1 $
  is shown in Figure 
\ref{energyflow} versus $\eta_{jet} $ compared to the
  predictions of  the two Monte Carlo
models. This ``pedestal''  energy  $E_{T,Ped}$ is calculated as
  follows. For every jet the transverse energy 
 is summed  in the region  $\Omega $  with area A defined by $-1 < \eta-\eta_{jet}
 < 1$
 and $-\pi < \phi - \phi_{jet} <  \pi$ around the analysed jet but excluding
 the jets
 themselves using a cone radius R=1.0.
 $E_{T,Ped}$ is finally given by:
  \begin{equation} E_{T,Ped} = \frac{1}{A} \sum_\Omega E_T
     \end{equation} 
The average transverse energy density measured outside of the jets (pedestal energy)
 is as high as 1.4 GeV at 
large $\eta_{jet}$ and can
therefore give, at small jet energies, a substantial contribution to the
transverse energy of a jet.
 A detailed study of the jet-jet  correlations using the
two Monte Carlo models shows good agreement between data and MC which justifies the use
 of these LO QCD Monte Carlo generators for the analysis.

\section{   Di-jet Cross-Section for \boldmath
 $E_{T,jet} >  4$ GeV}
Jets at the detector level are reconstructed using calorimeter clusters and
tracks. Monte Carlo events offer both the possibility to reconstruct jets at the
detector level  and to reconstruct jets using the
generated hadrons. The reconstructed jets are then used to calculate \xgamma
\ at the detector level (termed $x_{\gamma ,det}$) and at the hadron jet level
(termed  $x_{\gamma,jets}$).
 The correlation between these two quantities is used 
 to   unfold  \cite{unfold} the measured $x_{\gamma ,det}$ distribution to the hadron
jet level  in bins of $x_{\gamma,jets}$.
This correlation can be characterised by a Gaussian distribution in the
quantity  $(\log(x_{\gamma ,det}) - \log(x_{\gamma ,jets}))$ with a dispersion
of
$\sigma \approx 0.12$ and small non-Gaussian tails.
 Finally, the
differential cross-section  
$\ddiff\sigma / \ddiff\log(x_{\gamma,jets})$ is calculated.
After  reweighting the distributions according to  the new cross-section
derived during
 unfolding, both Monte Carlo simulations  describe  all aspects of
the measured data distributions in the detector equally well. 
\par
The dominant systematic  error (up to $24\%$ at low \xgamma) results 
from the uncertainty on
 the hadronic energy scale of  $ \pm 4\%$. 
The stability of the unfolding procedure is studied by starting with very
different cross-sections in the Monte Carlo simulation. This results in changes of the  unfolded cross-section of  less than 10\%.
The systematic errors of the   corrections for detector acceptance and
resolution are 
evaluated by using both Monte Carlo simulations (PYTHIA and PHOJET) and by using renormalisation
and factorisation scales  $0.5 \ \hat{p_t}$
in addition to the default choices  $\hat{p_t}$. The detector corrections  are found  to differ by
up to  $10\%$.
Additional experimental uncertainties arise from  the trigger efficiency
($5\%$) and the acceptance
of the small angle  electron detector
 and from the luminosity measurement (combined error $6\%$). All systematic errors are added in quadrature. 
\par

\begin{figure}[htb]
\epsfig{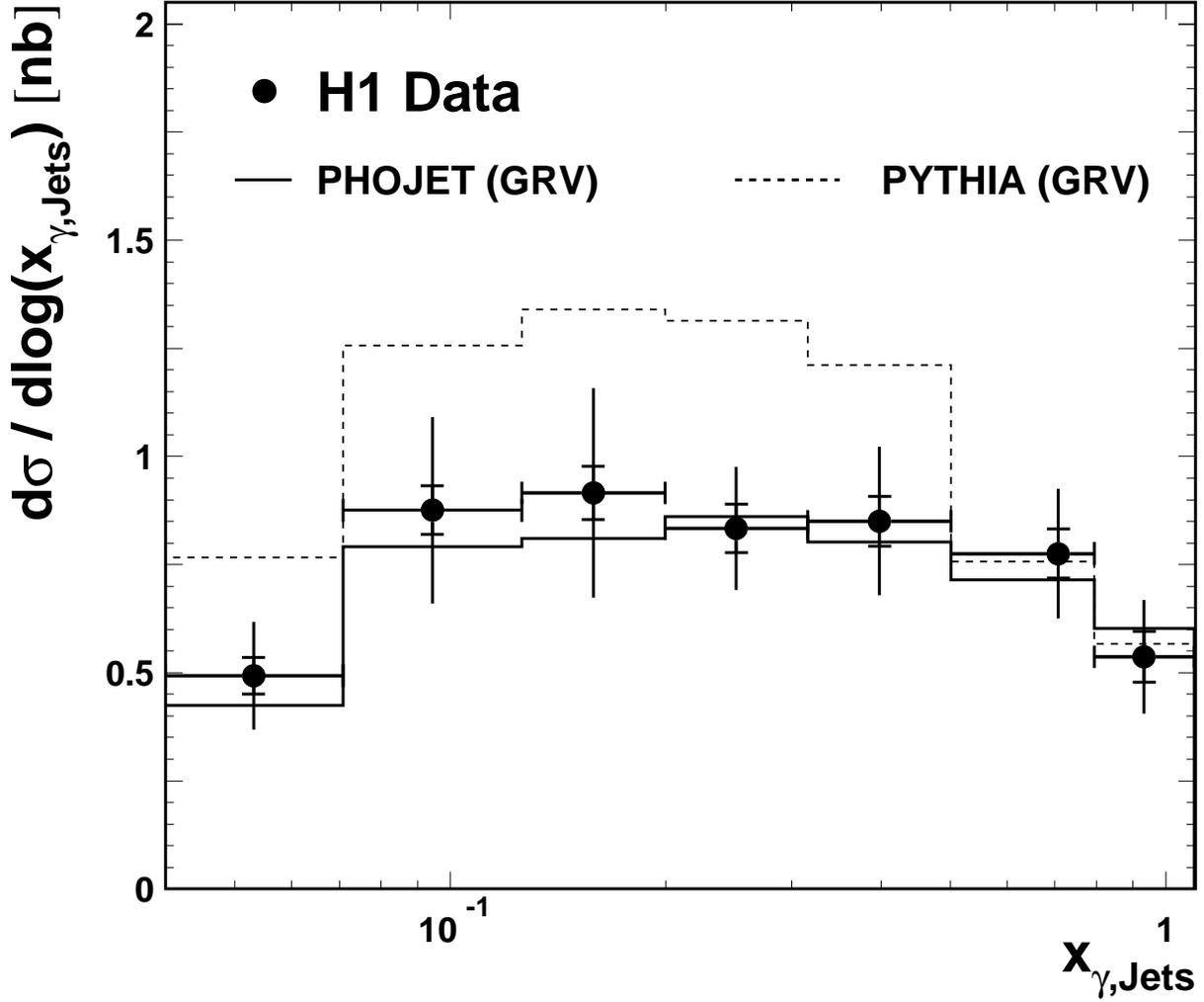}
\caption{The di-jet cross-section   at
  particle level (corrected for detector effects)
as a function of $x_{\gamma,jets}$ in the kinematic range 
$E_{T,jet} >  4 $ GeV,$ \ M_{1,2}  > 12 $ GeV,$ \ -0.5  <  \eta_{jet} <  2.5,
 \ |\eta_{jet1}-\eta_{jet2}|<1, \ 0.5  <  y_e  < 0.7$.  
 The jets were 
reconstructed with a cone algorithm using $ R = 0.7 $.
 The inner error bars give the statistical error and  the outer error
bars give the
total error.
 The data are 
compared to the predictions of PHOJET and PYTHIA using LO QCD matrix
elements, the GRV92-LO parton density
functions for  photon and proton \cite{grv} and $\hat{p_t}$ for the factorisation and
renormalisation scales.}

\label{dsigmahad}
\end{figure}

The di-jet cross-section  $\ddiff\sigma / \ddiff\log(x_{\gamma,jets})$  
is shown in Figure \ref{dsigmahad} and  Table \ref{hadtab} where the data
points are averages of  the results obtained using the two 
Monte Carlo simulations  for unfolding.
The measurement is made in the kinematic region
$E_{T,jet} > 4$ GeV, $M_{1,2} 
> 12$ GeV, $ -0.5 < \eta_{jet} < 2.5$, $ | \eta_{jet1} - \eta_{jet2} | < 1 $
and $0.5 < y_e < 0.7$.
The inner error bars reflect the statistical errors and the outer error bars
show the statistical and systematic errors added in quadrature.
This cross-section
determination  relies  on an
adequate   description of the energy flow and  of the angular correlations of
the jets,
both of which are achieved   by  the PYTHIA and PHOJET  Monte Carlo
simulations  as discussed in section
5. Remaining differences outside  the jet cones between data and either of
the Monte Carlo simulations  are comparable to the difference between the two
Monte Carlo simulations. 
The systematic error associated with the uncertainty in the description of the
energy flow, especially of the underlying event energy, is therefore
estimated as half the difference of the results obtained when
 unfolding with the
alternative Monte Carlo simulations. It amounts to 10 to 15 \%.

The absolute predictions of the PHOJET and PYTHIA models
using the same parton  density functions for the photon and the
proton and  the same 
factorisation and renormalisation scales are shown in Figure  \ref{dsigmahad}
 in comparison to the data.
 The two
predictions should be the same if this low $E_T$ jet sample were dominated
by
the effects of  hard scattering. However, they  differ
by almost a factor 2  for
$x_{\gamma} < 0.5$. This can be traced back to the 
parton transverse momentum spectra of the
selected di-jet events  which  differ greatly for PYTHIA and
PHOJET at low $\hat{ p_t}$ due to the different regularisation procedures.
PYTHIA predicts a much larger fraction of di-jet events
than PHOJET with parton $\hat{ p_t}$ between 2 and 4 GeV. Such partons 
produce jets with
 $E_{T,jet} > 4 $ GeV
because of a large underlying event energy in the jet cone.   
 We conclude that this event sample is strongly influenced
by effects such as the regularisation procedure and the underlying event
energy 
which makes  a  comparison to perturbative QCD predictions
difficult.
However, both models lead to a comparable good description of 
all aspects of the data once the predictions are reweighted to the 
measured $x_{\gamma}$  distribution. Therefore  the measured cross-section is 
a solid experimental result
which can be compared to every model which gives  a complete description of
hard $\gamma$p processes.




\section{ Analysis of Di-jet events for  \boldmath $E_{T,jet} > 6 $\ GeV}
The  determination of cross-sections usable for perturbative QCD analysis  requires a data
sample which is dominated by the effects of the  partons from the hard  scattering
process.  A more
restrictive data selection is therefore used for the subsequent analysis steps.

 The initial selection used here is as described in the previous section, but
 without the application of the jet-jet  mass cut, which becomes ineffective for the increased $E_T$ cut (see below). The following further
 requirements are then made:
\begin{enumerate}

\item   The transverse energy in the jet cone  for each jet is corrected for the
  average expected underlying event energy $E_{T,Ped}$  as function of 
 $\eta _{jet}$ ( Figure \ref{energyflow}). To do this, the average
 transverse energy density,  as determined outside  the jet cone using the  
Monte Carlo simulations,
  is  subtracted from the measured energy in the cone. Monte Carlo  studies show that
  this simple procedure leads to good agreement between the average  jet
   and  parton
 energies and at the same time improves the energy correlation 
  significantly for individual events.
  
After this subtraction the remaining transverse jet energy  has to be
larger than 6 GeV. This procedure reduces non-perturbative  effects much more
  effectively than just raising the
  cut on $E_{T,jet}$.
\item The pseudorapidity of each jet has to satisfy the requirement 
 $ \eta_{jet} > -0.9 - \ln[x_{\gamma,jets}]$. 
While this cuts hardly effects genuine di-jet events with 
 $E_{T,jet}>6$ GeV and  $| \eta_{jet1} - \eta_{jet2} | < 1$, it 
eliminates a large fraction of those events where one of the two jets used to 
reconstruct  $x_{\gamma}$ is not associated with the hard scattering process.
\end{enumerate} 

These additional cuts are introduced to achieve a good correlation
 between the measured
  and true   values of $x_{\gamma}$. 
They also reduce the differences in the parton distributions of the
  two Monte Carlo event samples. 
The selected di-jet sample contains  750 events.

\subsection{  Di-jet Cross-Section for 
  \boldmath $E_{T,jet} > 6$ GeV}

\begin{figure}[htb]
\epsfig{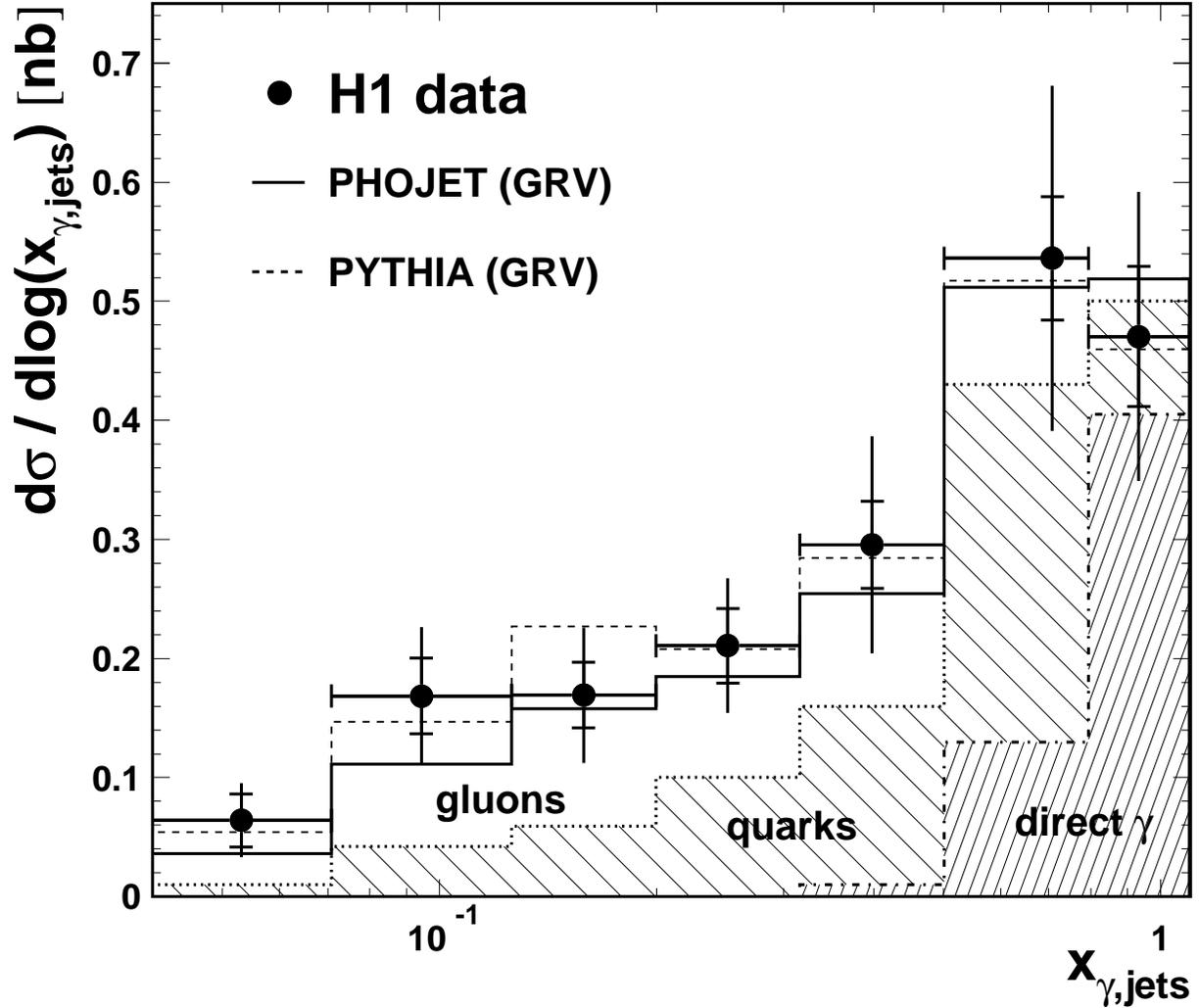}
\caption{The di-jet cross-section 
 at  particle level (corrected for detector effects)
as a function of  $x_{\gamma,jets}$ for the data in the kinematic range 
$E_{T,jet} >  6 $ GeV,$ \ -0.5  <  \eta_{jet}  <  2.5,
 \ |\eta_{jet1}-\eta_{jet2}| <  1, \ \eta_{jets}  >  -0.9  -  \ln(x_{\gamma,jets}), 
\ 0.5  <  y_e  <  0.7$. 
 The jets were 
reconstructed  with a cone algorithm using $ R = 0.7 $. The transverse energy of the 
jets was required to be greater than 6 GeV after subtraction of the average
pedestal 
energy.  The inner error bars give the statistical error and  the outer error
bars give the total error.
The data are 
compared to the LO QCD predictions of PHOJET and PYTHIA using the GRV92 LO parton density
functions for  photon and proton \cite{grv} and $\hat{p_t}$ for the factorisation and
renormalisation scale.  The 
contribution of direct  and resolved photon processes, as predicted by PHOJET, 
 are also shown.}

\label{crosssectionhighet}
\end{figure}

The cross-section $\ddiff\sigma / \ddiff\log(x_{\gamma,jets})$ 
is determined from the
event sample with  the jet transverse energy cut  $E_{T,jet} > 6 \ $ GeV
after pedestal energy subtraction.
 The result is shown in Figure~\ref{crosssectionhighet} and Table
 \ref{hadtab}  where the data
points are obtained by averaging  the results from the  two Monte Carlo 
used  for unfolding.
In Figure  \ref{crosssectionhighet} the data are compared to   the predictions of the two Monte Carlo simulations
using the GRV92 LO structure functions and  $\hat{p_t}$ for the renormalisation and
factorisation scales. For  PHOJET
 the 
contributions of resolved photon interactions due to  quarks and  gluons from
the photon and from direct photon interactions are  shown separately.  The
higher $E_T$ cut, combined with the pedestal subtraction,  strongly depopulates the
low \xgamma \ region and therefore also the region where gluons
 from the photon
dominate.  Nevertheless, the data remain  sensitive to the gluon distribution
down to \xgamma = 0.05. The differences between the PYTHIA and PHOJET predictions
based on the same parton densities  and using the same scales are 
now at a level between 10 and 40\%. 

\subsection{The  Effective Parton Distribution}
The di-jet cross-section in LO QCD is given by a sum of direct and resolved
photon contributions. The direct photon contribution depends only on the well
known 
parton distributions in the proton and can therefore be predicted.
The resolved part of the di-jet cross-section in LO QCD is a sum of quark-quark ($qq$), gluon-quark
($gq$) and
gluon-gluon ($gg$) scattering processes with different angular distributions
 and weights.
  To a good approximation,
 however, the differential
cross-section,   in the $\Delta \eta$ range chosen for this analysis,
 can be described with  a single effective subprocess using
effective parton distributions for the photon and proton and a
single differential parton-parton angular distribution
$\ddiff\hat{\sigma} / \ddiff \cos\hat{\Theta}$ \cite{combridge}.
 This is
because the angular distribution is very similar  for the largest
contributing  subprocesses within the present $\Delta \eta$ range. 
For resolved processes the differential cross-section can therefore be approximately  expressed as
$$ \frac {\ddiff^4\sigma ^{ep}}{\ddiff y \,\ddiff x_{\gamma} \,
 \ddiff x_p \,\ddiff \cos\hat{\Theta}} = \frac 1 {32\pi s_{ep}} 
\frac { f_{\gamma /e}}{y} \frac {f_{\gamma ,eff}(x_{\gamma})
f_{p,eff}(x_p)}{x_{\gamma}x_p}\frac {\ddiff\hat{\sigma}}{\ddiff \cos\hat{\Theta}} $$
Here the  effective parton distributions for the photon and proton can be written
$$f_{\gamma,eff}(x_{\gamma}) =   \left[q(x_{\gamma}) + \bar
{q}(x_{\gamma}) + 9/4 \ g(x_{\gamma}) \right]$$ 
$$f_{p,eff}(x_p) =   \left[q(x_p) + \bar{q}(x_p) + 9/4 \ g(x_p) \right]$$ 
$f_{\gamma /e}$ is the photon flux and $s_{ep}$ is the center of mass energy
squared of
 the $ep$ system.
The quark densities $q(x)$ comprise the sum over all flavours.
Since the parton densities in the proton are well constrained, the
 effective parton
density in the photon can be determined from the measured cross-section.
 Monte Carlo studies \cite{kaufmann} show that the  correlation  between $x_{\gamma ,det}$ 
 as reconstructed in the detector and  the  generated momentum fraction
$x_{\gamma}$
of the parton entering the hard scattering process from the photon side 
 for both PYTHIA and PHOJET is rather good. It can be characterised by a Gaussian
distribution of the  quantity
 $(\log(x_{\gamma ,jets})-\log(x_{\gamma}))$ with a dispersion  of $\sigma \approx 0.2$ and
only small non-Gaussian tails
 over the full measured 
 range of \xgamma.
 This correlation
is  used to correct  the measured jet cross-section for the effects of
hadronisation and  underlying event energy (which is only subtracted on average)
using the  unfolding method of   \cite{unfold}. 
During the unfolding procedure the direct and resolved photon contributions 
are  calculated keeping 
all parton densities in the proton  fixed to the
GRV92 LO parton distributions \cite{grv}, while
the effective parton density in the photon is adjusted
to get best agreement with the measured $x_{\gamma}$ distribution. This  determines the
effective parton density in the photon.

After unfolding and reweighting to the new effective parton density, all  Monte
Carlo
 distributions are in good agreement with the  data for
  both models \cite{kaufmann}. This
 demonstrates that it is possible in a leading order Monte Carlo model for the hard
scattering process  to get a good description of  the observed di-jet events
 when the optimised photon parton densities derived from our data are  used. 

\begin{figure}[htb]
\epsfig{clip=,file=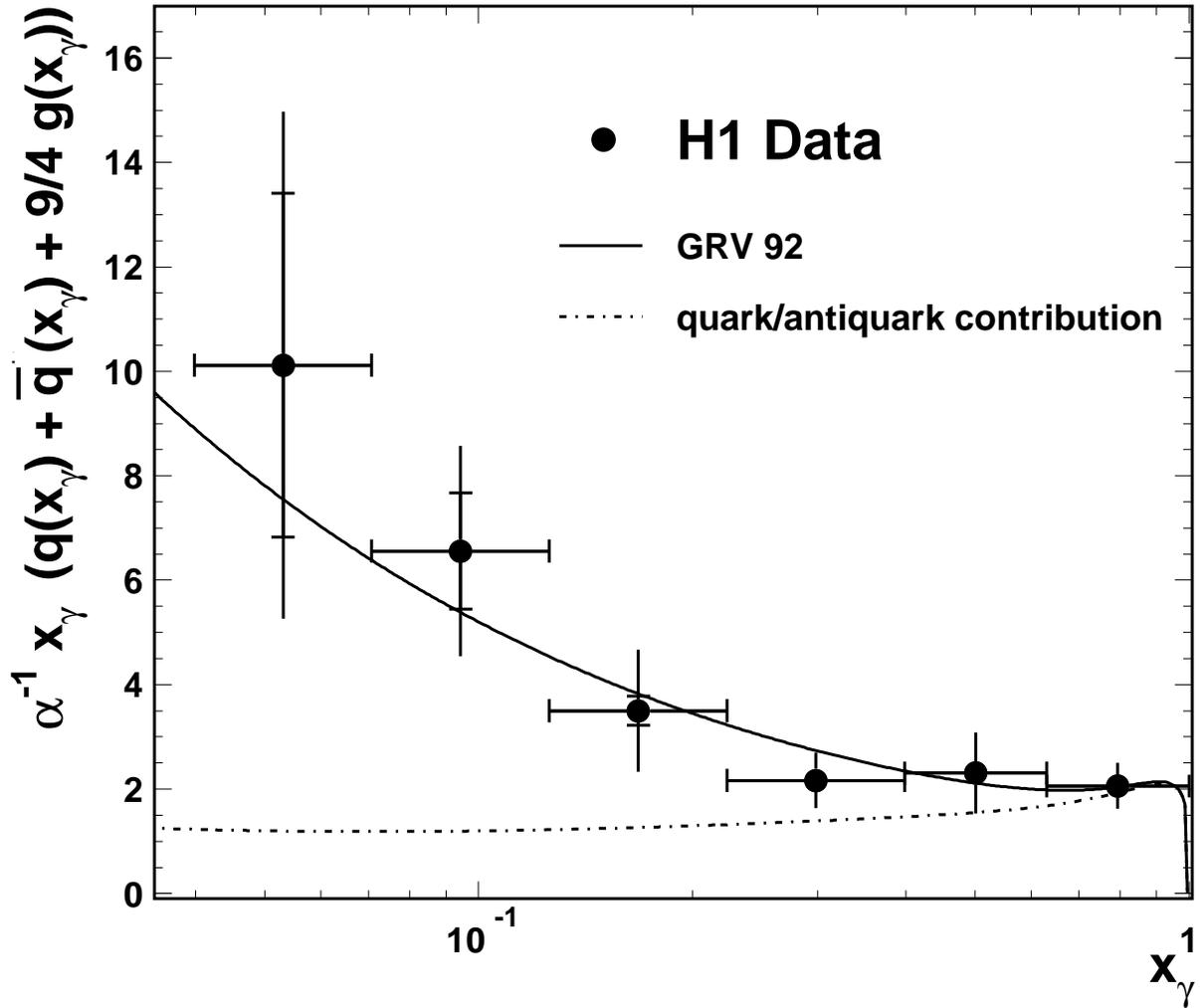,width=1.\linewidth}
\caption{Effective parton distribution
 $f_{\gamma,eff}=
  (\ q(x_{\gamma})\ +\bar {q} (x_{\gamma})  + \ 9/4 \ g(x_{\gamma})\ )$
 of the photon multiplied by $\alpha ^{-1}x_{\gamma}$ as a function of 
$x_{\gamma}$ for a mean scale $\hat{p_t}^2=74 $ GeV$^2$ of the hard partons.
The inner error bars give the statistical error and  the outer error
bars give the
total error.
The prediction based on the LO parametrisation of
   the parton distributions
 of the photon of  GRV92-LO \cite{grv} is also shown as well as  
 the sum of  quark and antiquark contributions (dash-dotted curve).}

\label{effpartons}

\end{figure}

 Figure \ref{effpartons} and Table \ref{parttab}  show the measured effective
 parton density of the photon multiplied by $\alpha ^{-1} x_{\gamma}$, where 
$\alpha $ is the fine structure constant. 
 The LO QCD expectation for the direct photon contribution, as given by the
 Monte Carlo simulations, has been subtracted. 
The magnitude and distribution  of this contribution can
be seen in Figure \ref{crosssectionhighet}. It
amounts to about 23\% of the selected di-jet  events. The measured points  
 correspond to an 
average scale  $ \hat{p_t}^2 = 74 $ GeV$^2$ as determined from the $\hat{p_t}$
of the partons in the weighted Monte Carlo sample which describes
 the data distributions.
Both statistical and total errors are given.  Systematic errors have been 
determined using the  method outlined in  section 6. A  systematic
error due to model uncertainties in the
$x_{\gamma}$  correlation is added. This error is dominated by the subtraction
of the underlying event energy and  taken to be  
 half the difference between  the  effective parton
distributions   using PYTHIA or PHOJET respectively for the unfolding. It 
 amounts to 15 to 20\%. The data points of
  Figure \ref{effpartons} and Table \ref{parttab} are finally obtained by
 averaging the  results obtained using  the two Monte Carlo simulations for unfolding.

The measured effective parton distribution  is compared to 
  the GRV92 LO parametrisation of
the parton densities in the photon which is obtained by a fit to $e^+e^-$
two-photon data alone \cite{grv}. These
 data constrain the quark density in the photon, but give
only indirect information on the gluon distribution via the
observed scaling violations.
The contribution of quarks plus antiquarks in the photon as given by the GRV92
  parametrisation is
  shown separately. It  includes the charm contribution (about 25\% of the
  quark  contribution) as calculated for $\gamma$p interactions.
 The predicted quark plus antiquark contribution describes the
 data   well at the highest values of \xgamma \  but falls far below  at small
 \xgamma. Within LO QCD the difference can only be attributed to a gluon contribution  which thus is shown to rise
 strongly towards low \xgamma. 
 \par
The extracted effective parton density constitutes the main result of this
analysis  at the
 parton level since, in contrast to the gluon density, it 
 can be extracted from our data alone.

\subsection{ The Gluon Distribution }
\begin{figure}[htb]
\epsfig{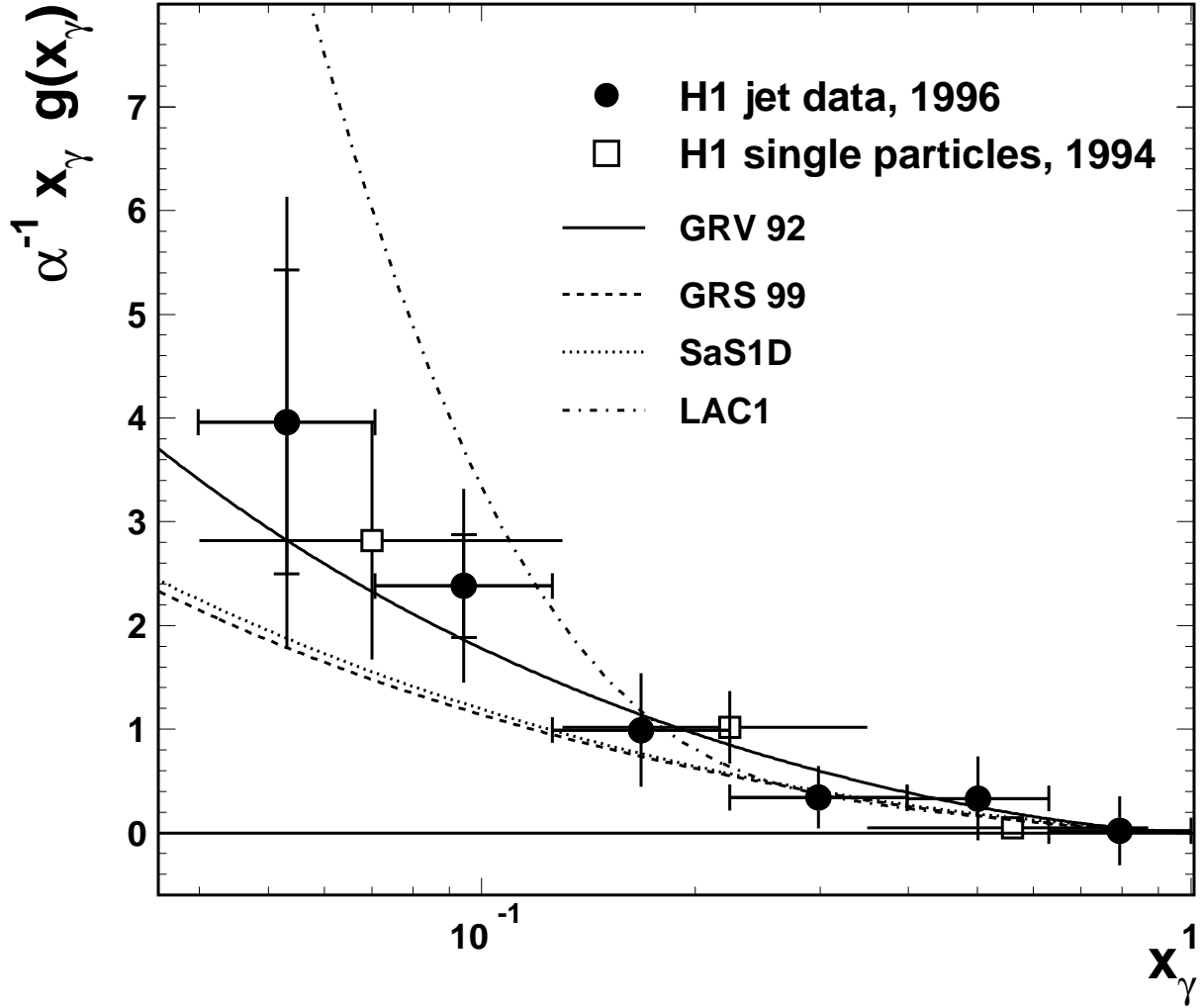}
\caption{Gluon distribution $ g(x_{\gamma})$ of the
  photon  multiplied by $\alpha ^{-1}x_{\gamma}$ as a function of $x_{\gamma}$
for a mean $\hat{p_t}^2=74 $ GeV$^2$ of the hard partons. The inner error
 bars give the statistical error and  the
 outer error bars give the
total error. The data points 
with open squares  show a previous
measurement of H1 which used single high $E_T$ particles to determine the LO 
gluon density of the photon \cite{highpttracks} at a mean scale $ \hat{p_t}^2
= 38 $ GeV$^2$.  The   LO parametrisations of the   gluon distribution
\cite{grv,lac1,GRS99,sas}
 based on fits to $e^+e^-$
two-photon data are also shown.}

\label{gluons}
\end{figure}

Since the quark  density in the photon is  well constrained by 
studies of photon-photon collisions in $e^+e^-$ data \cite{nisius} it
can be subtracted from the measured effective parton density within
 the present LO QCD
approach.  The subtraction is performed using the GRV92 LO parton
distributions which are in good agreement with the data and 
with other parametrisations. 
The 
resulting gluon distribution in the photon is shown in 
Figure \ref{gluons} and Table \ref{parttab}. The total error includes the
uncertainty
 of the  quark plus antiquark contribution
 in the photon 
 which is known with an error of less than 30\% for 
$0.1 < x_{\gamma} \ <0.8$ as derived directly from the measurement \cite{opal}.
This uncertainty increases up to 60\% at the  smallest
\xgamma
values considered here. This conservative error estimate 
covers also the uncertainty in the calculated charm contribution.

The gluon distribution is only large for  small \xgamma \ as expected. The
present 
measurement is in good agreement with an earlier H1 measurement
\cite{highpttracks} based on high transverse momentum  tracks in
photoproduction at scales $\hat{p_t}^2 = 38$ GeV$^2$ where systematic
uncertainties are very different. The analysis based on high transverse
momentum tracks has the advantage that it is hardly affected by the underlying
event energy. However, in such an  analysis it is not possible to define  a
quantity which is strongly correlated to \xgamma. This made the unfolding
procedure less effective and required coarse binning in \xgamma.

The measurements are compared 
to   LO parametrisations of the
gluon distribution.  Of the two older 
parametrisations, that  of GRV92 LO \cite{grv}  gives best
agreement with the data and has been used throughout this paper for comparisons
of the data with Monte Carlo predictions, whereas the parametrisation of LAC1
 \cite{lac1} shows a too steep rise at small \xgamma.
 The more recent parametrisations GRS99 \cite{GRS99}
and SaS1D \cite{sas} agree very well with each other but fall below the measured
distribution.
\section{Conclusions}

Two new measurements of the differential di-jet cross-section
$\ddiff\sigma /  \ddiff \log x_{\gamma ,jets}$ in  photoproduction at HERA are 
presented for rather low transverse jet energies. They reach  parton fractional energies 
down to $x_{\gamma ,jets} = 0.05 $,  a range where the gluons from the
photon are found to dominate the di-jet cross-section. This kinematic region is strongly affected by underlying
event energy and, for the  $E_{T,jet}> 4 $ GeV  selection, by the
 uncertainties in
the description of the transition from
hard to soft processes.
For di-jet events with $E_{T,jet} > 6 $ GeV, where the cut is applied 
after subtraction of the underlying event energy, the correlation to the parton dynamics is greatly improved.
Leading order QCD gives a good  description of these
data which makes possible a determination of the effective parton
 density in the
photon. This quantity 
is dominated by  the gluon density for $x_{\gamma} \le 0.2$ which is found to
rise
 strongly towards small
\xgamma. The result is in good agreement with earlier measurements of the H1 
collaboration but more precise.

\section*{Acknowledgements}

We are grateful to the HERA machine group whose outstanding
efforts have made and continue to make this experiment possible. 
We thank
the engineers and technicians for their work in constructing and now
maintaining the H1 detector, our funding agencies for 
financial support, the
DESY technical staff for continual assistance, 
and the DESY directorate for the
hospitality which they extend to the non DESY 
members of the collaboration.


\begin{table}[ht]
\renewcommand{\arraystretch}{1.5}
\begin{center}
\begin{small}
\begin{tabular}{|r@{}l|r@{}l|r@{}l|r@{}l|r@{}l|r@{}l|r@{}l|}

\hline
\multicolumn{2}{|c|}{}&
\multicolumn{6}{|c|}{$E_{T,jet} > \ 4 $ GeV}&
\multicolumn{6}{|c|}{$E_{T,jet} > \ 6 $ GeV} \\
\hline
\multicolumn{2}{|c|}{$x_{\gamma,jets}$}&
\multicolumn{2}{|c|}{$\frac{\ddiff\sigma}{\ddiff\log(x_{\gamma,jets})}[nb]$} & 
\multicolumn{2}{|c|}{stat.err.} & 
\multicolumn{2}{|c|}{total error}&
\multicolumn{2}{|c|}{$\frac{\ddiff\sigma}{\ddiff\log(x_{\gamma,jets})}[nb]$} & 
\multicolumn{2}{|c|}{stat.err.} & 
\multicolumn{2}{|c|}{total error}\\
\hline
0&.053& ~~~~~~0&.49& ~~~~0&.03& ~~~~~0&.14& ~~~~~~0&.06& ~~~~0&.02& ~~~~0&.03  \\
0&.094& ~~~~~~0&.88& ~~~~0&.04& ~~~~~0&.22& ~~~~~~0&.17& ~~~~0&.03& ~~~~0&.06  \\
0&.16
& ~~~~~~0&.92& ~~~~0&.04& ~~~~~0&.24& ~~~~~~0&.17& ~~~~0&.03& ~~~~0&.06  \\
0&.25& ~~~~~~0&.83& ~~~~0&.04& ~~~~~0&.14& ~~~~~~0&.21& ~~~~0&.03& ~~~~0&.06  \\
0&.40& ~~~~~~0&.85& ~~~~0&.04& ~~~~~0&.17& ~~~~~~0&.29& ~~~~0&.04& ~~~~0&.09  \\
0&.71& ~~~~~~0&.78& ~~~~0&.04& ~~~~~0&.14& ~~~~~~0&.54& ~~~~0&.05& ~~~~0&.14  \\
0&.93& ~~~~~~0&.54& ~~~~0&.04& ~~~~~0&.12& ~~~~~~0&.47& ~~~~0&.06& ~~~~0&.12   \\
\hline
\end{tabular}
\end{small}
\end{center}
\caption{The $\gamma$p di-jet cross-section 
$\ddiff\sigma / \ddiff\log(x_{\gamma ,jets})$ corrected to 
  hadron level with statistical and total error. Columns 2 to 4 give  the 
  cross-section  for 
the kinematic range  $E_{T,jet} > \ 4 $ GeV,
$  \  M_{1,2} \  > 12 $ GeV, $ -0.5 < \eta_{jets} < 2.5, \ |\eta_{jet1} - 
\eta_{jet2} | <
  1, \ 0.5 < y < 0.7$.
 The last three columns give the cross-section for
   $E_{T,jets} > \ 6 $ GeV 
  after pedestal energy subtraction in the kinematic range
 $  -0.5 < \eta_{jets} < 2.5, \ |\eta_{jet1} - \eta_{jet2} | < 1 \ , \ 0.5 < y < 0.7$, 
   $\eta_{jets}  >  -0.9  -  \ln(x_{\gamma,jets})$.  }

\label{hadtab}
\end{table}

\begin{table}[htb]
\renewcommand{\arraystretch}{1.5}
\begin{center}
\begin{tabular}{|r@{}l|r@{}l|r@{}l|r@{}l|r@{}l|r@{}l|r@{}l|}
\hline
\multicolumn{2}{|c|}{$x_{\gamma}$}&
\multicolumn{2}{|c|}{$\frac 1{\alpha} x_{\gamma} f_{\gamma,eff}(x_{\gamma})$} & 
\multicolumn{2}{|c|}{stat. error} &
\multicolumn{2}{|c|}{total error} &
\multicolumn{2}{|c|}{$\frac{1}{\alpha}x_{\gamma}g_{\gamma}(x_{\gamma})$} &
\multicolumn{2}{|c|}{stat. error} & 
\multicolumn{2}{|c|}{total error}\\

\hline
0&.053& ~~~~~10&.1& ~~~~~3&.3& ~~~~~4&.9& ~~~~~4&.0& ~~~~~1&.4&  ~~~~~2&.1\\
0&.094& ~~~~~~6&.6& ~~~~~1&.1& ~~~~~2&.0& ~~~~~2&.4& ~~~~~0&.5& ~~~~~0&.9 \\
0&.17& ~~~~~~3&.5& ~~~~~0&.3& ~~~~~1&.2& ~~~~~0&.99& ~~~~~0&.12&  ~~~~~0&.55\\
0&.30& ~~~~~~2&.2& ~~~~~0&.1& ~~~~~0&.5& ~~~~~0&.34& ~~~~~0&.03&  ~~~~~0&.30 \\
0&.50& ~~~~~~2&.3& ~~~~~0&.2& ~~~~~0&.8& ~~~~~0&.33& ~~~~~0&.08& ~~~~~0&.40 \\
0&.79& ~~~~~~2&.1& ~~~~~0&.1& ~~~~~0&.4& ~~~~~0&.02& ~~~~~0&.03&  ~~~~~0&.33\\
\hline
\end{tabular}
\end{center}
\caption{ Columns 2 to 4: The effective parton density
 $f_{\gamma,eff}=
  (\ q(x_{\gamma})\ +\bar {q} (x_{\gamma})  + \ 9/4 \ g(x_{\gamma})\ )$. Last
 three columns: The  gluon density
 $  g(x_{\gamma})$. Both measurements are given 
 for an average scale 
 $\hat{p_t}^2 = 74$  GeV$^2$. Statistical and
 total errors are given separately.}
\label{parttab}
\end{table}


\begin{thebibliography}{99}
\footnotesize{
\bibitem{h1_tracks} H1 Collab., I. Abt {\it et al.}, \Journal{\PLB}{328}{1994}{176}.
\bibitem{h1_old}H1 Collab., T. Ahmed {\it et al.},
  \Journal{\NPB}{445}{1995}{195}.
\bibitem{h1.jet98}  H1 Collab., C. Adloff {\it et al.},
  \Journal{\EUR}{1}{1998}{97}.
\bibitem{zeus_old} ZEUS Collab., M. Derrick {\it et al.},
  \Journal{\PLB}{384}{1996}{401}; \\
 ZEUS Collab., J. Breitweg  {\it et al.}, \Journal{\EUR}{4}{1998}{591}.
\bibitem{jet_hera} H1 Collab., S. Aid {\it et al.},
  \Journal{\ZPC}{70}{1996}{17}.
\bibitem{ZEUSNLO} ZEUS Collab., J. Breitweg {\it et al.}, \Journal{\EUR}{11}{1999}{35}.
\bibitem{h1detector} H1 Collab., I. Abt {\it et al.},
  \Journal{\NIMA}{386}{1997}{310 and 348}.
\bibitem{highq**2}H1 Collab., C. Adloff {\it et al.}, 
``Measurement of Neutral and Charged Current Cross-Sections in Positron-Proton
Collisions at Large Momentum Transfer'',
\it DESY-99-107, hep-ex/9908059, \rm accepted by 
 \it  Eur. Phys. J. C. \rm
\bibitem{cdfcone} CDF Collab., F. Abe {\it et al.}, \Journal{\PRD}{45}{1992}{1448}.
\bibitem{phojet} PHOJET Monte Carlo,  R. Engel, \Journal{\ZPC}{66}{1995}{203}.
\bibitem{pythia} T. Sj\"ostrand, CERN-TH-6488 (1992), \it
  Comput. Phys. Commun.\rm  \bf 82 \rm (1994) 74.
\bibitem{jetset} T. Sj\"ostrand, M. Bengtsson,\ \it  Comput. Phys. Commun. \rm \bf 43
  \rm (1987) 367.
\bibitem{grv} M. Gl\"uck, E. Reya, A. Vogt, \Journal{\PRD}{46}{1992}{1973};\\
M. Gl\"uck, E. Reya, A. Vogt, \Journal{\ZPC}{53}{1992}{127}.

\bibitem{kaufmann} O. Kaufmann,\ \it  PhD Thesis Universit\"at Heidelberg \rm
  1999, unpublished.\\
http://www-h1.desy.de/psfiles/theses/h1th-185.ps
\bibitem{unfold} G. D'Agostini, \Journal{\NIMA}{362}{1995}{487}.
\bibitem{combridge}B.L. Combridge and C.J. Maxwell,
  \Journal{\NPB}{239}{1984}{429}.
\bibitem{nisius} R. Nisius in proceedings of PHOTON99, {\em
    Nucl. Phys. Proc. Suppl.}{\bf B  48}\ {(2000)}.
\bibitem{opal} OPAL Collab., K. Ackerstaff {\it et al.}, \Journal {\PLB}{412}{1997}{225}.
\bibitem{highpttracks}H1 Collab., C. Adloff {\it et. al.}, \Journal{\EUR} {10}{1999}{363}.}
\bibitem{lac1} H. Abramowicz, K. Charchula, A. Levy,
  \Journal{\PLB}{269}{1991}{458}.
\bibitem{GRS99} M. Gl\"uck, E. Reya, I. Schienbein, \Journal{\PRD}{60}{1999}{54019}.
\bibitem{sas} G. A. Sch\"uler, T. Sj\"ostrand, \Journal{\PLB}{376}{1996}{193}.

\end{thebibliography}
\end{document}